\documentclass[a4paper,11pt]{article}
\usepackage{pos}

\usepackage{xcolor}

\newcommand{\civ}{\ifmmode {\rm C}\,{\sc iv} \else C\,{\sc iv}\fi}
\newcommand{\CIV}{\ifmmode {\rm C}\,{\sc iv}\,\lambda1549 \else 
	           C\,{\sc iv}\,$\lambda1549$\fi}
\newcommand{\oiii}{O\,{\sc iii}}

\newcommand{\Lo}{L_{\rm UV}}

\newcommand{\Lx}{L_{\rm X}}

\title{The relation between X-ray and UV emission in quasars}

\author*[a]{Susanna Bisogni}

\affiliation[a]{INAF-IASF Milano,\\
  Via Alfonso Corti 12, 20133, Milano, Italy}

\emailAdd{susanna.bisogni@inaf.it}

\abstract{The correlation between the X-ray and UV luminosities observed in quasars, spanning a wide redshift range and holding true for several decades in both spectral bands, suggests the presence of a universal mechanism governing the transfer of energy from the accretion disc to the hot corona.
In this study, we leverage X-ray spectroscopic data extracted from the Chandra Source Catalog 2.0 for a sample of over $2000$ quasars from the Sloan Digital Sky Survey Data Release 14 (SDSS DR14). Our analysis reveals a reduced intrinsic dispersion in the $\Lx-\Lo$ relation at higher redshifts ($\delta < 0.2$ dex) compared to previous studies relying on photometric data from catalogs. Additionally, our findings confirm the stability of this relation up to redshifts of approximately $4.5$.
The $\Lx-\Lo$ relation can also serve as a tool to investigate the physics of accretion by identifying outliers—sources that exhibit a different state of the accretion disc/hot corona system compared to the average population. For instance, X-ray-weak quasars are sources with reduced X-ray emissions due to a radiatively inefficient state of the corona, and their optical properties suggest the presence of a powerful accretion disc wind.
The wealth of spectroscopic data available in the CSC 2.0/SDSS catalogs opens up the opportunity for a more comprehensive exploration of the central engine in AGN.}

\FullConference{
Multifrequency Behaviour of High Energy Cosmic Sources XIV (MULTIF2023)\\
12-17 June 2023\\
Palermo, Italy\\}

\begin{document}
\maketitle

\section{Introduction}

The relationship between X-ray and UV emissions in quasars has been a subject of study since the late seventies \citep{Tananbaum1979, Zamorani1981}. The prevailing interpretation of energy production in AGN involves the comptonisation of UV seed photons, emitted by the accretion disc, to the X-rays by the electrons in the hot-corona{\citep{HaardtMaraschi1991, HaardtMaraschi1993}. This \emph{two-phase model}, however, does not explain the observed non-linear slope of the relation \citep[$\gamma \sim 0.6$, e.g.][]{Vignali2003, Strateva2005, Steffen2006, Just2007, Lusso2010, Young2010}, that implies that more luminous sources in the UV band are relatively less luminous in the X-rays. Additionally, the model fails to account for the persistence of the X-ray emission, given the fast cooling times implied by comptonisation, in the absence of a sustaining and refueling energy process.

Despite the unknowns on the underlying physics, the non-linearity observed in the $\Lx-\Lo$ relation has enabled the inference of the luminosity distance for quasars \citep{RisalitiLusso2015, RisalitiLusso2019, Lusso2020}. As a result, this relation has been employed as a tool for cosmological applications, provided that its dispersion is sufficiently small to allow for precise distances measurements.
In the last years, significant efforts have focused on demonstrating that the \emph{observed} dispersion in the relation is primarily influenced by observational factors, particularly the calibration of X-ray measurements \citep{RisalitiLusso2015, LussoRisaliti2016, LussoRisaliti2017, Bisogni2017d, Salvestrini2019, RisalitiLusso2019, Bisogni2021, Sacchi2022}. By implementing effective selection criteria, that exclude sources with UV and X-ray fluxes not representative of the intrinsic emission from the accretion disc and hot-corona\footnote{This can be accounted for both by observational problems, like the one connected to calibration in the X-rays, and by obscuration/contamination of the intrinsic emission. A notable example of this last case are Radio Loud (RL) and Broad Absorption Line (BAL) quasars, the first ones characterised by an additional contribution in the X-rays due to the jets and the second ones by strong absorption in the UV, but the selection criteria exclude any kind of source absorbed or contaminated in either band by emission from other components and the host galaxy.}, we can approach the \emph{intrinsic} dispersion. 

Besides being necessary to enable cosmological applications, the tightness of the relation over several decades in luminosities in both bands, along with the stability of the slope over cosmic time (e.g \citep{Bisogni2021}), also conveys important information about the universality of the physical mechanism that couples the accretion disc and the corona, that must hold for luminous and fainter engines.
On a more general basis, the $\Lx-\Lo$ relation describes the physics of the accretion in quasars. Therefore, it is crucial to understand how the gravitational energy lost by the matter accreting onto the SuperMassive Black Hole (SMBH) is converted into light, and, if present, into winds and jets. All of these represent  ways in which the central engine impacts its surrounding environment on much larger scales than the scale of accretion, ultimately influencing the evolution of the host galaxies \citep[e.g. ][]{Harrison2014, Fiore2017, Cicone2018}).
The nebular regions surrounding the central engine, Broad Line Region (BLR) and Narrow Line Region (NLR), respond to the ionisation caused by the primary emission from the central engine by emitting broad and narrow lines in the UV and optical range. Photons in the soft X-ray/ far UV range, originating from the hot-corona and the inner part of the accretion disc, are responsible for the production of the high ionisation potential lines, such as the \civ\ in the UV and the [\oiii] in the optical.
It is clear, therefore, that in order to delve into the physics underlying this relation and analyse the impact of the energy produced by the central engine on the surrounding nebular regions, which ultimately influences scales up to kiloparsecs, \emph{spectroscopic} information is vital.

Several studies have examined quasars' spectra in both X-ray and UV bands, often focusing on small samples that are representative of specific sub-populations of quasars, defined by either redshift or luminosity range \citep[e.g ][]{Nardini2019, Zappacosta2020, Lusso2021, Laurenti2022, Trefoloni2023}. However, these samples may not fully capture the diversity of the entire quasar population.
To explore the connection between the accretion scale and the kpc scale for the entire quasar population, we require a \emph{statistically significant} sample. The sample presented in this work, for the first time, combines a high level of statistical significance and spectroscopic information in both the X-ray and UV bands.

\section{A statistically significant spectroscopic sample in the X-ray band}

All the studies that aimed to reduce the dispersion in the relation by using statistically significant samples relied on photometric information in the X-ray band \citep{RisalitiLusso2015, LussoRisaliti2016, LussoRisaliti2017, Bisogni2017d}. For the first time, the Chandra Source Catalog 2.0 \citep[CSC 2.0, ][]{Evans2010} enabled us to retrieve spectroscopic information in the X-ray band for thousands of sources \citep{Bisogni2021}.
This sample was obtained by cross-matching a pre-selection of the SDSS DR14\footnote{We excluded BAL and RL quasars and selected sources that are not dust-absorbed or host galaxy contaminated in the UV/optical band.} with the CSC 2.0, resulting in more than 3000 spectroscopic data products ready to be used for scientific analysis.
Fig. \ref{fig1} shows the $\Lx-\Lo$ relation\footnote{The rest frame flux at 2 keV, and consequently the $\Lx$, was measured by fitting the X-ray spectrum with \emph{Xspec}: we used a power law corrected for Galactic absorption and a \emph{cflux (calculate flux)} component as a model. This allowed us to retrieve the intrinsic flux at 2 keV as one of the free parameters of the fit, together with the slope and the normalisation of the power law. The rest frame flux at 2500 \AA\ (and then $\Lo$) was instead computed from interpolation of the SED of the source by using the multi-wavelength photometric data available from the UV to the Near Infra Red bands. Detailed information on the analysis can be found in \cite{Bisogni2021}.} for the final sample in \cite{Bisogni2021}, which combines the CSC 2.0 dataset with \emph{Chandra} COSMOS Legacy data \citep{Civano2016, Marchesi2016}. The final selection was obtained after applying the filtering criteria in the X-rays, mainly aimed at excluding absorbed sources and at cleaning the sample for the Eddington bias, i.e. the inclusion of sources because of a positive fluctuation with respect to their average emission.
It spans the remarkable redshift range of $\sim 0.5 - 4.5$, especially noteworthy given that these data are entirely from catalogs.

\begin{figure}
\begin{center}
	\includegraphics[width=0.7\columnwidth]{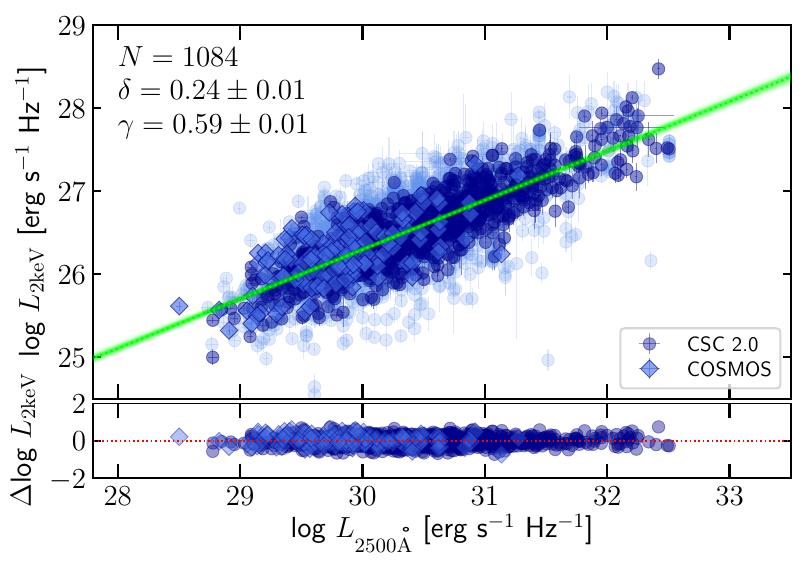}
\caption{The $\Lx-\Lo$ relation for the final sample published in \cite{Bisogni2021}, combining CSC 2.0/SDSS DR14 and COSMOS Legacy data. Sources in the parent sample that have been excluded by the selection criteria are shown in light colors.}
    \label{fig1}
    \end{center}
\end{figure}

\begin{figure}
\begin{center}
	\includegraphics[width=0.7\columnwidth]{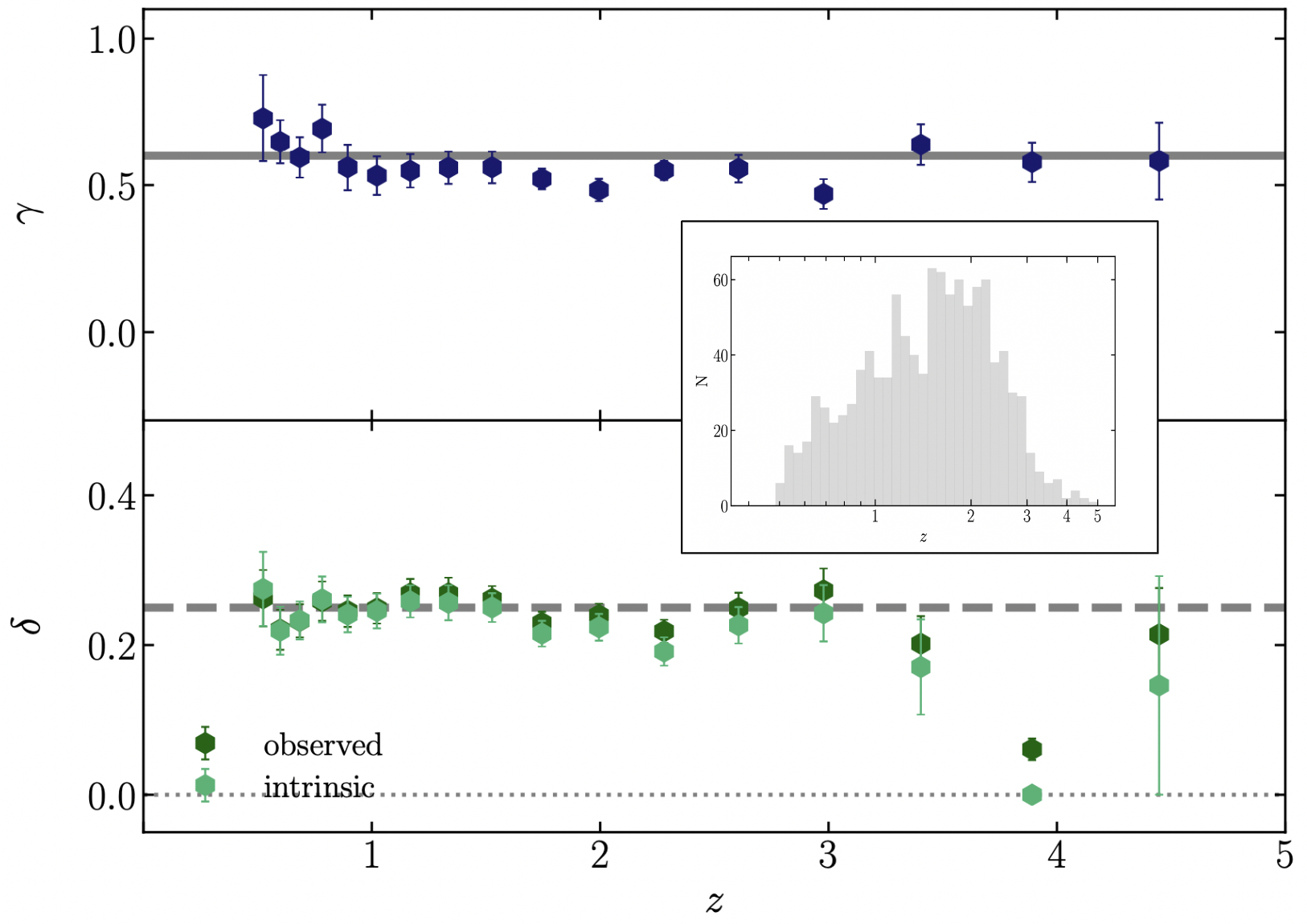}
\caption{Analysis of the $f_{\rm{X}}-f_{\rm{UV}}$ relation in bins of redshift adapted from \cite{Bisogni2021}. \emph{Top panel:} Slope $\gamma$ as a function of redshift. The slope is stable around the value of $0.6$ (grey solid line) over the redshift range spanned by the sample ($z \sim 0.48-4.5$, the redshift distribution is shown in the inset).
\emph{Bottom panel:} Dispersion $\delta$ as a function of redshift. The dispersion remains around the median value $\sim 0.25$ dex (grey dashed line). The values show a decreasing trend with redshift due to the fact that, as the redshift increases, the probability also increases that the X-ray flux, whose measurement errors contribute the most to the dispersion, originates from targeted observations of the source, while at lower redshifts, the bulk of the data comes from serendipitous observations, which are more susceptible to calibration errors. In the highest redshift bins, the average value for the dispersion is $\delta \sim 0.15$ dex.
Both the intrinsic dispersion (derived from the emcee regression analysis) and the observed dispersion are displayed.}
    \label{fig2}
    \end{center}
\end{figure}

The high statistics available (>1500 sources) allowed us to split the sample in redshift bins and explore the behaviour of slope ($\gamma$) and dispersion ($\delta$) of the relation across cosmic time. The redshift bins are chosen to be small enough that the differences in the distances within the same bin are negligible when compared to the observed dispersion in the relation. This allows for using the fluxes in place of the luminosities and for studying the evolution of slope and dispersion in a cosmologically independent way.

Our analysis shows that the slope $\gamma$ remains stable up to redshift $z\sim4.5$ (Fig. \ref{fig2}, top panel).
The analysis of the (non-)evolution of the relation with redshift is crucial for both cosmological and physical studies. A stable slope over cosmic time is essential for using quasars as standard candles. Additionally, it serves as a strong indication of the universality of the physical mechanism governing the coupling of the accretion disc and hot corona.
 
The accuracy achieved through the spectroscopic analysis in measuring the flux at $2$ keV, the proxy for the emission of the hot-corona, led to a dispersion $\delta \sim 0.15$ dex at the highest redshift bins (Fig. \ref{fig2}, bottom panel).
This level of accuracy is comparable to that of samples with dedicated X-ray observations \citep{Lusso2020}, which is very likely the case here as well (the observations in the CSC 2.0 at these redshifts were obtained through dedicated observations).


\section{The $\Lx-\Lo$ relation as a touchstone for the state of accretion of quasars}

We have confirmed a few key aspects of the $\Lx-\Lo$ relation: the emission from the corona increases at a slower rate than that from the disc in more luminous sources. Furthermore, the relation remains stable over several decades in luminosity in both bands and, also, across cosmic time.  Several works investigated the interplay between accretion disc and hot-corona, many invoking a coupling via magnetic fields for explaining the X-ray emission in quasars \citep[e.g., ][]{MerloniFabian2001, Liu2002, Merloni2003, Cheng2020}, others clumpy accretion flows \citep[e.g., ][]{IshibashiCourvoisier2009} or modified viscosity prescriptions in the accretion disc depending on the accretion status of the source \citep{Arcodia2019}. Other works examined the differences between the emissions coming from disc and corona in different accretion regimes \citep{Liu2021}.

Despite all these efforts, we still lack a clear comprehension of the coupling between the two innermost components and a clear explanation of the establishment of the $\Lx-\Lo$ relation.
A more comprehensive approach would be examining the entire $\Lx-\Lo$ plane, rather than just the relation itself, in order to characterise the accretion status of quasars based on their X-ray and UV properties, regardless of whether they conform to the relation or not.
In this regard, our focus has shifted from selecting the sample to reduce the dispersion and identify sources with a ``canonical" coupling of accretion disc and hot corona, i.e. lying on the relation. Instead, we are now interested in examining the position of the sources of the entire quasar population in the $\Lx-\Lo$ plane, without any selection applied except to ensure that the sample consists exclusively of blue quasars, thereby excluding dust-reddened or host galaxy-contaminated sources.

In this context, the $\Lx-\Lo$ relation can be thought as a reference for the ``canonical'' coupling of accretion disc and corona and be used as a landmark in the $\Lx-\Lo$ plane describing the accretion of quasars.

Given the $\Lx-\Lo$ relation, for any $\Lo$ we can infer an ``expected" $\Lx$, or equivalently, an ``expected" optical to X-ray spectral index $\alpha_{\rm{OX}}= - \rm{log}$ $(\Lx / \Lo)$ $/\rm{log} (\nu_{\rm{X}}/\nu_{\rm{UV}})$ \citep{Tananbaum1979}, describing the steepness of the SED between the X and UV bands. If we want to measure the position of a source in the $\Lx-\Lo$ plane, i.e. the distance of a source from the locus of the $\Lx-\Lo$ relation, we can do that through the parameter $\Delta \alpha_{\rm{OX}}= \alpha_{\rm{OX}_{exp}} - \alpha_{\rm{OX}_{obs}} = - 0.384 * \rm{log} ($ $L_{\rm{X}_{exp}}$ $/ L_{\rm{X}_{obs}})$.


\subsection{The X-ray Weak population}

The approach examining the entire $\Lx-\Lo$ plane rather than only the locus of the relation was first motivated by the identification of a class of objects, selected to be part of the blue quasar population, but exhibiting a markedly different behaviuor, as found in a previous study \citep{Nardini2019}.

As part of the project using quasars for measuring cosmological distances, we selected a sample of sources from the SDSS DR7 at $z \sim 3$  with the goal of populating the Hubble diagram at high redshift. The sources were specifically selected to be blue and unabsorbed in the UV band and, by design, they were expected to lie on the $\Lx-\Lo$ relation.

The \emph{XMM-Newton} observations carried out for these sources (cycle 16, proposal ID: 080395, PI: G. Risaliti) revealed that roughly a third of them exhibit significantly lower X-ray fluxes than anticipated based on their UV emission assuming the $\Lx-\Lo$ relation (they lie within 2 and 3 $\sigma$ below the $\Lx-\Lo$ relation). Additionally, the X-ray spectra, which are flatter on average compared to their normal counterparts, do not show any sign of absorption in the X-ray band \citep{Nardini2019}.
The explanation proposed for the behaviour of these \emph{intrinsically X-ray Weak} sources is a different state of the corona with respect to the radiatively efficient state of the sources that follow the relation \citep{Nardini2019, Lusso2021}. 

Considering the size of the parent sample (30 quasars), the fraction of sources falling into the X-ray weak sub-population is large, ranging between $25\%$ and $30\%$, depending on the threshold adopted to distinguish between the \emph{X-ray Weak} and the \emph{X-ray Normal}, i.e. quasars with a X-ray emission consistent with the expectations of the $\Lx-\Lo$ relation.
Such a high fraction of X-ray Weak quasars has never been reported in samples of radio quiet, non-BAL quasars (e.g \cite{Pu2020}), but has been confirmed by other recent studies on samples of similarly highly accreting quasars \citep{Zappacosta2020, Laurenti2022}, finding up to 40\% of X-ray Weak sources.

While, by design of the sample selection, X-ray Weak and Normal sources share similar UV luminosity, differences in the profiles of the broad lines emitted hint at the possible presence of a nuclear wind in the X-ray Weak population, which may deplete the reservoir of UV photons from the accretion disc, potentially leading to the starvation of the corona in this class of objects \citep{Nardini2019, Lusso2021}.

\subsection{A radiatively inefficient corona: the effect on nebular regions}

\begin{figure}
\begin{center}
	\includegraphics[width=0.7\columnwidth]{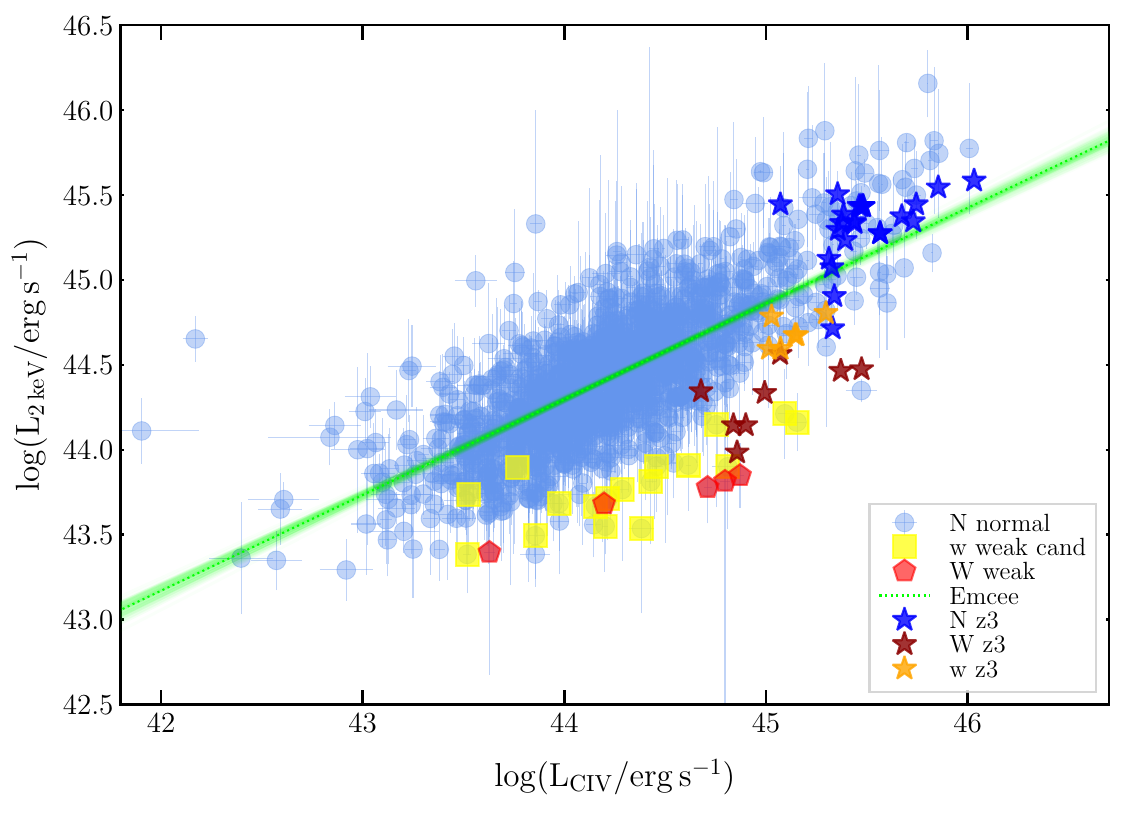}
\caption{$\Lx-L_{\rm{CIV}}$ relation for the CSC 2.0/SDSS DR14 sample. Red pentagons, yellow squares and light blue circles represent X-ray Weak sources (W weak, $\Delta \alpha_{OX}<-0.3$), X-ray Weak candidates (w weak cand, $-0.3<\Delta \alpha_{OX}<-0.2$) and X-ray Normal sources (N normal, $-0.2<\Delta \alpha_{OX}$) respectively. Stars indicate sources from the \emph{XMM} $z\sim3$ sample \citep{Nardini2019, Lusso2021} (red = X-ray Weak (W z3), yellow = X-ray Weak candidates (w z3), blue = X-ray Normal (N z3). The emcee regression line is based on the CSC 2.0/SDSS DR14 data points, while the \emph{XMM} sample is included for comparison.}
    \label{fig3}
    \end{center}
\end{figure}

The reduced availability of seed photons from the accretion disc and the resulting depletion in the hot corona emission are expected to have an effect on the nebular regions surrounding the central engine. 
In particular, the more energetic photons emitted by the hot corona, whose energy range spans the far UV and soft X-rays bands, are responsible for the production of the high ionisation potential lines emitted by the BLR and NLR.

The connection between the X-rays emission and the BLR emission is evident in the case of one of the most prominent emission features in the UV spectra of quasars, \civ\ (ionisation potential 47.9 eV): the integrated luminosity of the line is tightly correlated with the luminosity at 2 keV \citep{Lusso2021}.
This result can be interpreted in the context of the $\Lx-\Lo$ relation. At high UV luminosity, where $\alpha_{OX}$ steepens compared to values observed in quasars characterised by lower UV luminosity, corresponding to a larger difference between the UV and the X-ray luminosities, there is a deficit of ionising photons in the far-UV/soft X-rays. This deficit, in turn, impacts the production of high ionisation potential lines, such as \civ. 
However, it remains unclear why X-ray weak sources exhibit an excess of \civ\ compared to the X-ray normal, average population. This seems to be in apparent contrast with the global trend observed in the context of the Eigenvector 1 \citep{BorosonGreen1992, Wang1996}, where weak \civ\ emissions are associated with sources having relatively weaker X-ray emission \citep{Richards2011}. 
One possible explanation for this is provided by the combination of the line excitation mechanisms for \civ, dominated by collisions, whose rate strongly depends on the temperature of the gas and therefore on the amount of X-ray photons available, and the non linearity of the $\Lx-\Lo$ relation, which implies that for X-ray Normal sources with similar X-ray luminosity, but lower UV luminosity, the number of ionising photons is lower \citep{Timlin2021, Lusso2021}.

\begin{figure}
\begin{center}
	\includegraphics[width=0.7\columnwidth]{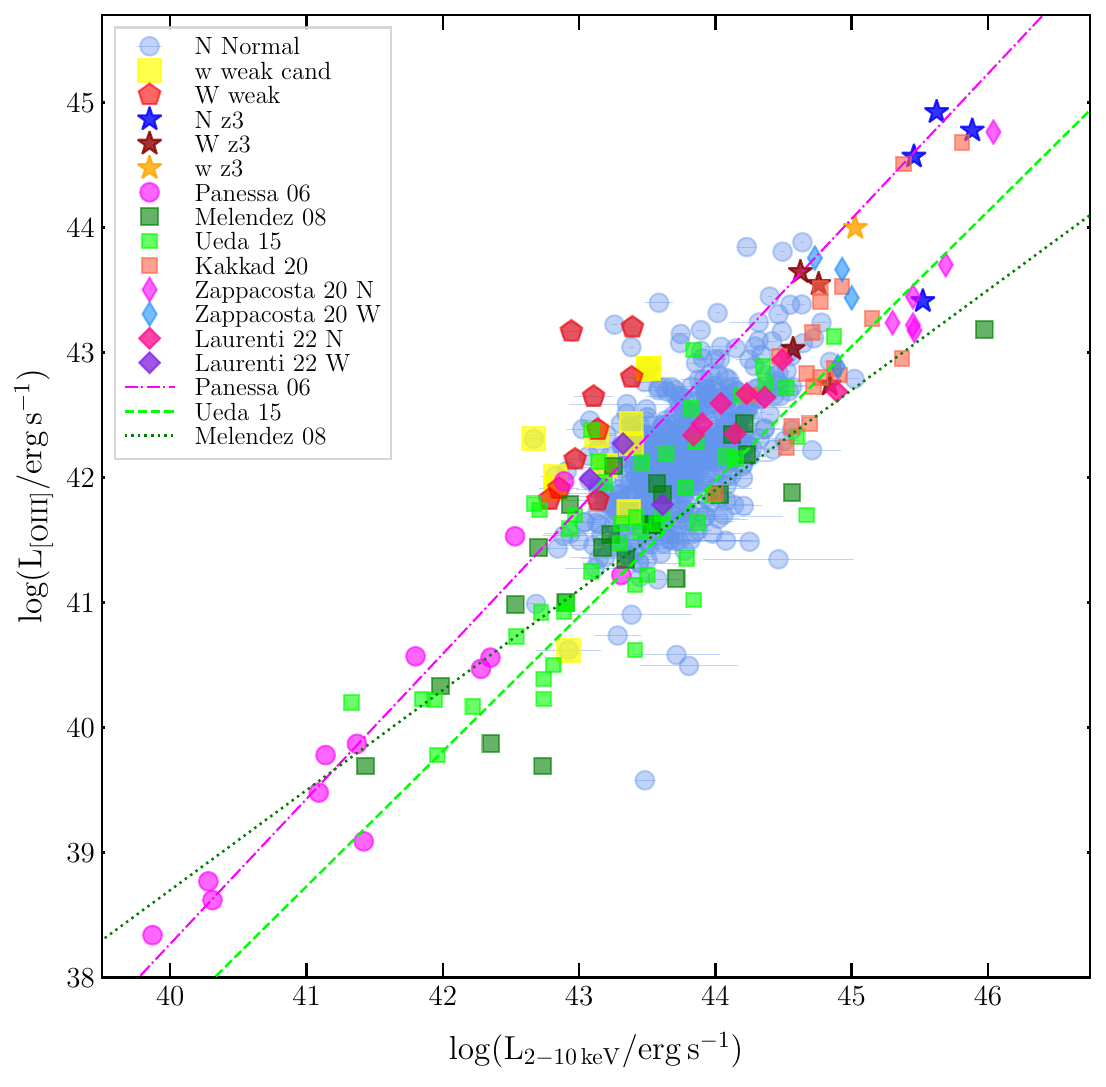}
\caption{$L_{2-10 \rm{keV}}-L_{\rm{[OIII]}}$ relation for the CSC 2.0/SDSS DR14 sample compared to the samples presented in \cite{Trefoloni2023} (\emph{XMM} $z \sim 3$ sample, stars (red = X-ray Weak (W z3), yellow = X-ray Weak candidates (w z3), blue = X-ray Normal (N z3)) and references therein. For CSC 2.0/SDSS DR14 data, red pentagons, yellow squares and light blue circles represent X-ray Weak sources (W weak), X-ray Weak candidates (w weak cand) and X-ray Normal (N normal) sources respectively. The luminosities of the [\oiii] line in the different studies have been computed from the line after continuum subtraction and have been corrected for both Galactic and NLR extinction in \cite{Panessa2006, Ueda2015}, only for NLR extinction in \cite{Zappacosta2020}, only for Galactic extinction in \cite{Laurenti2022, Trefoloni2023} and this work, and neither for Galactic nor NLR extinction in \citep{Melendez2008, Kakkad2020}.}
    \label{fig4}
    \end{center}
\end{figure}

\begin{figure}
\begin{center}
	\includegraphics[width=0.7\columnwidth]{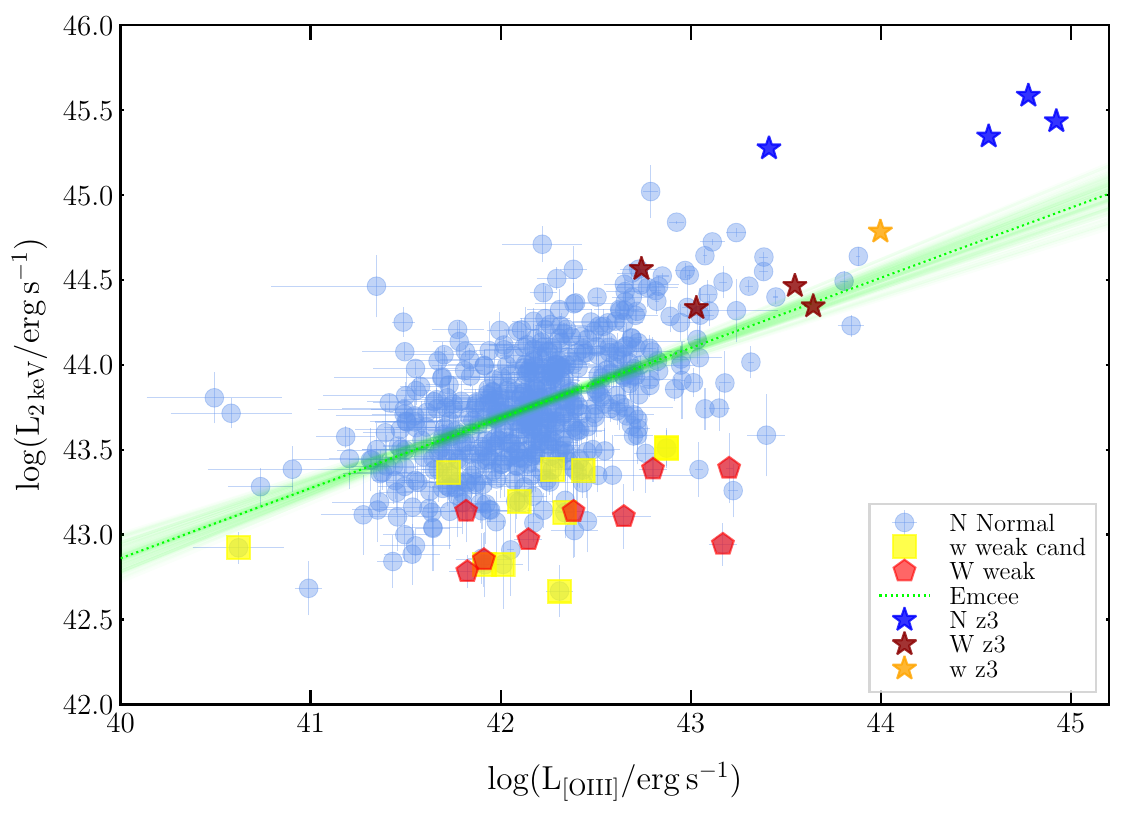}
\caption{$\Lx-L_{\rm{[OIII]}}$ relation for the CSC 2.0/SDSS DR14 sample. Red pentagons, yellow squares and light blue circles represent X-ray Weak sources (W weak), X-ray Weak candidates (w weak cand) and X-ray Normal (N normal) sources respectively. Stars indicate sources from the \emph{XMM} $z \sim 3$ sample. (red = X-ray Weak (W z3), yellow = X-ray Weak candidates (w z3), blue = X-ray Normal (N z3)). The emcee regression line is based on the CSC 2.0/SDSS DR14 data points, while the XMM sample is included for comparison.}
    \label{fig5}
    \end{center}
\end{figure}

The CSC 2.0 sample allows us to broaden our investigation to include the entire population of quasars with spectroscopic information in the X-ray band, rather than just focusing on the bright end or high redshift sources. Fig. \ref{fig3} shows the $\Lx-L_{\rm{CIV}}$ relation for the CSC 2.0-SDSS DR14 sample (circles) and for the $z \sim 3$ \emph{XMM-Newton} sample of \cite{Nardini2019, Lusso2021} (stars).
\civ\ is sampled here in the redshift range $1.2 < z < 4.5$\footnote{the lower limit is imposed by SDSS wavelength coverage, while the upper limit by the redshift range sampled by the CSC 2.0 sample}.
X-ray Weak and X-ray Weak ``candidates" are defined, following \cite{Nardini2019}, by selecting sources with a $\Delta \alpha_{OX}$, i.e. the difference between the observed and expected $\alpha_{OX}$\footnote{The expected $\alpha_{OX}$ is computed following the $\Lx-\Lo$ relation in \cite{Bisogni2021}}, $<-0.3$ (red) and $-0.3< \Delta \alpha_{OX} <-0.2$ (yellow) respectively.
X-ray Weak quasars confirm the behaviour shown in the $z \sim 3$ sample, i.e. an excess of \civ\ with respect to the bulk of quasar population.

Thanks to the redshift range spanned by the CSC 2.0 sample, we can extend the study to the connection between the X-ray emission and the NLR by examining the [\oiii] line (ionisation potential 35.1 eV), available for sources at $0.48 < z < 0.9$\footnote{The lower limit is imposed by the selection of the parent sample (sources with $z < 0.48$ are likely contaminated by host-galaxy emission), while the upper limit is set by SDSS wavelength coverage}.
The presence of a correlation between [\oiii] and the X-ray emission is not unexpected \citep{Mulchaey1994, Heckman2005, Panessa2006, Melendez2008, Ueda2015}, considering [\oiii] is used as a proxy of the bolometric luminosity in AGN \citep[e.g][]{Pennell2017}, which is basically dominated by the UV emission, and that the X-ray emission, as a first approximation, serves as a proxy for the bolometric luminosity as well \citep[e.g][]{Runnoe2012}, even though the non-linearity of the $\Lx-\Lo$ relation adds some complexity.

To compare our sample with others with information in the [\oiii] and X-ray band \citep[][ and references therein]{Trefoloni2023}, we plot in Fig. \ref{fig4} the relation between the luminosity of the line and the X-ray luminosity in the $2-10$ keV band.
Our sample as a whole aligns well with the relationship when compared to the others. However, in contrast to the \emph{XMM} sample at $z\sim 3$\citep{Trefoloni2023}, where, despite the weakness of [\oiii] profiles in X-ray weak sources, this subclass does not deviate from the overall population, here we observe that the X-ray Weak and X-ray Weak candidates populate a specific region of space relative to the relationship, showing again an excess of  line flux with respect to X-ray Normal sources with analogous X-ray emission.
Fig. \ref{fig5} shows the same behaviour, this time in the relation between the [\oiii] luminosity and the monochromatic luminosity at 2keV, where X-ray Weak and Weak candidates are located below the main relation, as in the case of the \civ\ line.
Despite the differences between the two nebular regions in terms of distance from the central engine and density, this seems to suggest that the starvation of the hot corona might also affect high ionisation lines emitted by the NLR at the kpc scale, and not just those emitted by the BLR.

\section{Future perspective: the $\Lx-\Lo$ plane as a function of $\Delta \alpha_{OX}$}
The $\Lx-\Lo$ relation can serve as a benchmark and can be employed to discriminate between X-ray Weak and X-ray Normal sources. 

The natural extension of this work is to widen our approach and explore the possibility of using the entire $\Lx-\Lo$ plane as a diagnostic tool for quasar accretion.
This can be achieved by investigating variations in UV and X-ray properties based on their ``distance" from the $\Lx-\Lo$ relation, quantified as $\Delta \alpha_{OX}$. In this context, X-ray Weak and Weak candidate sources represent the most distant objects,  characterised by $\Delta \alpha_{OX}<0.3$ and $<0.3\Delta \alpha_{OX}<0.2$, respectively. 
The final goal is to take a complete census of the quasar population and to describe the state of the accretion as a function of the position on the $\Lx - \Lo$ plane.
We are tackling the problem using two different methods. First, we are studying the properties measured on X-ray and UV spectra, such as fluxes, equivalent widths, and line properties, to determine if some of these properties change continuously with respect to $\Delta \alpha_{OX}$. Second, we are comparing UV and X-ray stacked spectra for different bins of $\Delta \alpha_{OX}$.
This will be the subject of a forthcoming publication.

%

\bibliography{bisogni}

\providecommand{\href}[2]{#2}\begingroup\raggedright\begin{thebibliography}{10}

\bibitem{Tananbaum1979}
H.~{Tananbaum}, Y.~{Avni}, G.~{Branduardi}, M.~{Elvis}, G.~{Fabbiano},
  E.~{Feigelson} et~al., \emph{{X-ray studies of quasars with the Einstein
  Observatory}}, \href{https://doi.org/10.1086/183100}{\emph{\apjl} {\bfseries
  234} (1979) L9}.

\bibitem{Zamorani1981}
G.~{Zamorani}, J.P.~{Henry}, T.~{Maccacaro}, H.~{Tananbaum}, A.~{Soltan},
  Y.~{Avni} et~al., \emph{{X-ray studies of quasars with the Einstein
  Observatory. II}}, \href{https://doi.org/10.1086/158815}{\emph{\apj}
  {\bfseries 245} (1981) 357}.

\bibitem{HaardtMaraschi1991}
F.~{Haardt} and L.~{Maraschi}, \emph{{A two-phase model for the X-ray emission
  from Seyfert galaxies}}, \href{https://doi.org/10.1086/186171}{\emph{\apjl}
  {\bfseries 380} (1991) L51}.

\bibitem{HaardtMaraschi1993}
F.~{Haardt} and L.~{Maraschi}, \emph{{X-ray spectra from two-phase accretion
  disks}}, \href{https://doi.org/10.1086/173020}{\emph{\apj} {\bfseries 413}
  (1993) 507}.

\bibitem{Vignali2003}
C.~{Vignali}, W.N.~{Brandt} and D.P.~{Schneider}, \emph{{X-Ray Emission from
  Radio-Quiet Quasars in the Sloan Digital Sky Survey Early Data Release: The
  {\ensuremath{\alpha}}$_{ox}$ Dependence upon Ultraviolet Luminosity}},
  \href{https://doi.org/10.1086/345973}{\emph{\aj} {\bfseries 125} (2003) 433}
  [\href{https://arxiv.org/abs/astro-ph/0211125}{{\ttfamily
  astro-ph/0211125}}].

\bibitem{Strateva2005}
I.V.~{Strateva}, W.N.~{Brandt}, D.P.~{Schneider}, D.G.~{Vanden Berk} and
  C.~{Vignali}, \emph{{Soft X-Ray and Ultraviolet Emission Relations in
  Optically Selected AGN Samples}},
  \href{https://doi.org/10.1086/431247}{\emph{\aj} {\bfseries 130} (2005) 387}
  [\href{https://arxiv.org/abs/astro-ph/0503009}{{\ttfamily
  astro-ph/0503009}}].

\bibitem{Steffen2006}
A.T.~{Steffen}, I.~{Strateva}, W.N.~{Brandt}, D.M.~{Alexander},
  A.M.~{Koekemoer}, B.D.~{Lehmer} et~al., \emph{{The X-Ray-to-Optical
  Properties of Optically Selected Active Galaxies over Wide Luminosity and
  Redshift Ranges}}, \href{https://doi.org/10.1086/503627}{\emph{\aj}
  {\bfseries 131} (2006) 2826}
  [\href{https://arxiv.org/abs/astro-ph/0602407}{{\ttfamily
  astro-ph/0602407}}].

\bibitem{Just2007}
D.W.~{Just}, W.N.~{Brandt}, O.~{Shemmer}, A.T.~{Steffen}, D.P.~{Schneider},
  G.~{Chartas} et~al., \emph{{The X-Ray Properties of the Most Luminous Quasars
  from the Sloan Digital Sky Survey}},
  \href{https://doi.org/10.1086/519990}{\emph{\apj} {\bfseries 665} (2007)
  1004} [\href{https://arxiv.org/abs/0705.3059}{{\ttfamily 0705.3059}}].

\bibitem{Lusso2010}
E.~{Lusso}, A.~{Comastri}, C.~{Vignali}, G.~{Zamorani}, M.~{Brusa}, R.~{Gilli}
  et~al., \emph{{The X-ray to optical-UV luminosity ratio of X-ray selected
  type 1 AGN in XMM-COSMOS}},
  \href{https://doi.org/10.1051/0004-6361/200913298}{\emph{\aap} {\bfseries
  512} (2010) A34} [\href{https://arxiv.org/abs/0912.4166}{{\ttfamily
  0912.4166}}].

\bibitem{Young2010}
M.~{Young}, M.~{Elvis} and G.~{Risaliti}, \emph{{The X-ray Energy Dependence of
  the Relation Between Optical and X-ray Emission in Quasars}},
  \href{https://doi.org/10.1088/0004-637X/708/2/1388}{\emph{\apj} {\bfseries
  708} (2010) 1388} [\href{https://arxiv.org/abs/0911.0474}{{\ttfamily
  0911.0474}}].

\bibitem{RisalitiLusso2015}
G.~{Risaliti} and E.~{Lusso}, \emph{{A Hubble Diagram for Quasars}},
  \href{https://doi.org/10.1088/0004-637X/815/1/33}{\emph{\apj} {\bfseries 815}
  (2015) 33} [\href{https://arxiv.org/abs/1505.07118}{{\ttfamily 1505.07118}}].

\bibitem{RisalitiLusso2019}
G.~{Risaliti} and E.~{Lusso}, \emph{{Cosmological Constraints from the Hubble
  Diagram of Quasars at High Redshifts}},
  \href{https://doi.org/10.1038/s41550-018-0657-z}{\emph{Nature Astronomy}
  {\bfseries 3} (2019) 272} [\href{https://arxiv.org/abs/1811.02590}{{\ttfamily
  1811.02590}}].

\bibitem{Lusso2020}
E.~{Lusso}, G.~{Risaliti}, E.~{Nardini}, G.~{Bargiacchi}, M.~{Benetti},
  S.~{Bisogni} et~al., \emph{{Quasars as standard candles. III. Validation of a
  new sample for cosmological studies}},
  \href{https://doi.org/10.1051/0004-6361/202038899}{\emph{\aap} {\bfseries
  642} (2020) A150} [\href{https://arxiv.org/abs/2008.08586}{{\ttfamily
  2008.08586}}].

\bibitem{LussoRisaliti2016}
E.~{Lusso} and G.~{Risaliti}, \emph{{The Tight Relation between X-Ray and
  Ultraviolet Luminosity of Quasars}},
  \href{https://doi.org/10.3847/0004-637X/819/2/154}{\emph{\apj} {\bfseries
  819} (2016) 154} [\href{https://arxiv.org/abs/1602.01090}{{\ttfamily
  1602.01090}}].

\bibitem{LussoRisaliti2017}
E.~{Lusso} and G.~{Risaliti}, \emph{{Quasars as standard candles. I. The
  physical relation between disc and coronal emission}},
  \href{https://doi.org/10.1051/0004-6361/201630079}{\emph{\aap} {\bfseries
  602} (2017) A79} [\href{https://arxiv.org/abs/1703.05299}{{\ttfamily
  1703.05299}}].

\bibitem{Bisogni2017d}
S.~{Bisogni}, G.~{Risaliti} and E.~{Lusso}, \emph{{A Hubble Diagram for
  Quasars}}, \href{https://doi.org/10.3389/fspas.2017.00068}{\emph{Frontiers in
  Astronomy and Space Sciences} {\bfseries 4} (2017) 68}
  [\href{https://arxiv.org/abs/1712.07515}{{\ttfamily 1712.07515}}].

\bibitem{Salvestrini2019}
F.~{Salvestrini}, G.~{Risaliti}, S.~{Bisogni}, E.~{Lusso} and C.~{Vignali},
  \emph{{Quasars as standard candles II. The non-linear relation between UV and
  X-ray emission at high redshifts}},
  \href{https://doi.org/10.1051/0004-6361/201935491}{\emph{\aap} {\bfseries
  631} (2019) A120}.

\bibitem{Bisogni2021}
S.~{Bisogni}, E.~{Lusso}, F.~{Civano}, E.~{Nardini}, G.~{Risaliti}, M.~{Elvis}
  et~al., \emph{{The Chandra view of the relation between X-ray and UV emission
  in quasars}}, \href{https://doi.org/10.1051/0004-6361/202140852}{\emph{\aap}
  {\bfseries 655} (2021) A109}
  [\href{https://arxiv.org/abs/2109.03252}{{\ttfamily 2109.03252}}].

\bibitem{Sacchi2022}
A.~{Sacchi}, G.~{Risaliti}, M.~{Signorini}, E.~{Lusso}, E.~{Nardini},
  G.~{Bargiacchi} et~al., \emph{{Quasars as high-redshift standard candles}},
  \href{https://doi.org/10.1051/0004-6361/202243411}{\emph{\aap} {\bfseries
  663} (2022) L7} [\href{https://arxiv.org/abs/2206.13528}{{\ttfamily
  2206.13528}}].

\bibitem{Harrison2014}
C.M.~{Harrison}, D.M.~{Alexander}, J.R.~{Mullaney} and A.M.~{Swinbank},
  \emph{{Kiloparsec-scale outflows are prevalent among luminous AGN: outflows
  and feedback in the context of the overall AGN population}},
  \href{https://doi.org/10.1093/mnras/stu515}{\emph{\mnras} {\bfseries 441}
  (2014) 3306} [\href{https://arxiv.org/abs/1403.3086}{{\ttfamily 1403.3086}}].

\bibitem{Fiore2017}
F.~{Fiore}, C.~{Feruglio}, F.~{Shankar}, M.~{Bischetti}, A.~{Bongiorno},
  M.~{Brusa} et~al., \emph{{AGN wind scaling relations and the co-evolution of
  black holes and galaxies}},
  \href{https://doi.org/10.1051/0004-6361/201629478}{\emph{\aap} {\bfseries
  601} (2017) A143} [\href{https://arxiv.org/abs/1702.04507}{{\ttfamily
  1702.04507}}].

\bibitem{Cicone2018}
C.~{Cicone}, M.~{Brusa}, C.~{Ramos Almeida}, G.~{Cresci}, B.~{Husemann} and
  V.~{Mainieri}, \emph{{The largely unconstrained multiphase nature of outflows
  in AGN host galaxies}},
  \href{https://doi.org/10.1038/s41550-018-0406-3}{\emph{Nature Astronomy}
  {\bfseries 2} (2018) 176} [\href{https://arxiv.org/abs/1802.10308}{{\ttfamily
  1802.10308}}].

\bibitem{Nardini2019}
E.~{Nardini}, E.~{Lusso}, G.~{Risaliti}, S.~{Bisogni}, F.~{Civano}, M.~{Elvis}
  et~al., \emph{{The most luminous blue quasars at 3.0 < z < 3.3. I. A tale of
  two X-ray populations}},
  \href{https://doi.org/10.1051/0004-6361/201936911}{\emph{\aap} {\bfseries
  632} (2019) A109} [\href{https://arxiv.org/abs/1910.04604}{{\ttfamily
  1910.04604}}].

\bibitem{Zappacosta2020}
L.~{Zappacosta}, E.~{Piconcelli}, M.~{Giustini}, G.~{Vietri}, F.~{Duras},
  G.~{Miniutti} et~al., \emph{{The WISSH quasars project. VII. The impact of
  extreme radiative field in the accretion disc and X-ray corona interplay}},
  \href{https://doi.org/10.1051/0004-6361/201937292}{\emph{\aap} {\bfseries
  635} (2020) L5} [\href{https://arxiv.org/abs/2002.00957}{{\ttfamily
  2002.00957}}].

\bibitem{Lusso2021}
E.~{Lusso}, E.~{Nardini}, S.~{Bisogni}, G.~{Risaliti}, R.~{Gilli},
  G.T.~{Richards} et~al., \emph{{The most luminous blue quasars at 3.0 < z <
  3.3. II. C IV/X-ray emission and accretion disc physics}},
  \href{https://doi.org/10.1051/0004-6361/202141356}{\emph{\aap} {\bfseries
  653} (2021) A158} [\href{https://arxiv.org/abs/2107.02806}{{\ttfamily
  2107.02806}}].

\bibitem{Laurenti2022}
M.~{Laurenti}, E.~{Piconcelli}, L.~{Zappacosta}, F.~{Tombesi}, C.~{Vignali},
  S.~{Bianchi} et~al., \emph{{X-ray spectroscopic survey of highly accreting
  AGN}}, \href{https://doi.org/10.1051/0004-6361/202141829}{\emph{\aap}
  {\bfseries 657} (2022) A57}
  [\href{https://arxiv.org/abs/2110.06939}{{\ttfamily 2110.06939}}].

\bibitem{Trefoloni2023}
B.~{Trefoloni}, E.~{Lusso}, E.~{Nardini}, G.~{Risaliti}, G.~{Bargiacchi},
  S.~{Bisogni} et~al., \emph{{The most luminous blue quasars at 3.0 < z < 3.3.
  III. LBT spectra and accretion parameters}},
  \href{https://doi.org/10.1051/0004-6361/202346024}{\emph{\aap} {\bfseries
  677} (2023) A111} [\href{https://arxiv.org/abs/2305.07699}{{\ttfamily
  2305.07699}}].

\bibitem{Evans2010}
I.N.~{Evans}, F.A.~{Primini}, K.J.~{Glotfelty}, C.S.~{Anderson},
  N.R.~{Bonaventura}, J.C.~{Chen} et~al., \emph{{The Chandra Source Catalog}},
  \href{https://doi.org/10.1088/0067-0049/189/1/37}{\emph{\apjs} {\bfseries
  189} (2010) 37} [\href{https://arxiv.org/abs/1005.4665}{{\ttfamily
  1005.4665}}].

\bibitem{Civano2016}
F.~{Civano}, S.~{Marchesi}, A.~{Comastri}, M.C.~{Urry}, M.~{Elvis},
  N.~{Cappelluti} et~al., \emph{{The Chandra Cosmos Legacy Survey: Overview and
  Point Source Catalog}},
  \href{https://doi.org/10.3847/0004-637X/819/1/62}{\emph{\apj} {\bfseries 819}
  (2016) 62} [\href{https://arxiv.org/abs/1601.00941}{{\ttfamily 1601.00941}}].

\bibitem{Marchesi2016}
S.~{Marchesi}, F.~{Civano}, M.~{Elvis}, M.~{Salvato}, M.~{Brusa}, A.~{Comastri}
  et~al., \emph{{The Chandra COSMOS Legacy survey: optical/IR
  identifications}},
  \href{https://doi.org/10.3847/0004-637X/817/1/34}{\emph{\apj} {\bfseries 817}
  (2016) 34} [\href{https://arxiv.org/abs/1512.01105}{{\ttfamily 1512.01105}}].

\bibitem{MerloniFabian2001}
A.~{Merloni} and A.C.~{Fabian}, \emph{{Accretion disc coronae as magnetic
  reservoirs}},
  \href{https://doi.org/10.1046/j.1365-8711.2001.04060.x}{\emph{\mnras}
  {\bfseries 321} (2001) 549}
  [\href{https://arxiv.org/abs/astro-ph/0009498}{{\ttfamily
  astro-ph/0009498}}].

\bibitem{Liu2002}
B.F.~{Liu}, S.~{Mineshige} and K.~{Shibata}, \emph{{A Simple Model for a
  Magnetic Reconnection-heated Corona}},
  \href{https://doi.org/10.1086/341877}{\emph{\apjl} {\bfseries 572} (2002)
  L173} [\href{https://arxiv.org/abs/astro-ph/0205257}{{\ttfamily
  astro-ph/0205257}}].

\bibitem{Merloni2003}
A.~{Merloni}, \emph{{Beyond the standard accretion disc model: coupled magnetic
  disc-corona solutions with a physically motivated viscosity law}},
  \href{https://doi.org/10.1046/j.1365-8711.2003.06496.x}{\emph{\mnras}
  {\bfseries 341} (2003) 1051}
  [\href{https://arxiv.org/abs/astro-ph/0302074}{{\ttfamily
  astro-ph/0302074}}].

\bibitem{Cheng2020}
H.~{Cheng}, B.F.~{Liu}, J.~{Liu}, Z.~{Liu}, E.~{Qiao} and W.~{Yuan},
  \emph{{Magnetic-reconnection-heated corona in active galactic nuclei: refined
  disc-corona model and application to broad-band radiation}},
  \href{https://doi.org/10.1093/mnras/staa1250}{\emph{\mnras} {\bfseries 495}
  (2020) 1158} [\href{https://arxiv.org/abs/2006.02665}{{\ttfamily
  2006.02665}}].

\bibitem{IshibashiCourvoisier2009}
W.~{Ishibashi} and T.J.L.~{Courvoisier}, \emph{{AGN UV and X-ray luminosities
  in clumpy accretion flows}},
  \href{https://doi.org/10.1051/0004-6361:200810043}{\emph{\aap} {\bfseries
  495} (2009) 113} [\href{https://arxiv.org/abs/0812.0473}{{\ttfamily
  0812.0473}}].

\bibitem{Arcodia2019}
R.~{Arcodia}, A.~{Merloni}, K.~{Nandra} and G.~{Ponti}, \emph{{Testing the
  disk-corona interplay in radiatively-efficient broad-line AGN}},
  \href{https://doi.org/10.1051/0004-6361/201935874}{\emph{\aap} {\bfseries
  628} (2019) A135} [\href{https://arxiv.org/abs/1907.10069}{{\ttfamily
  1907.10069}}].

\bibitem{Liu2021}
H.~{Liu}, B.~{Luo}, W.N.~{Brandt}, M.S.~{Brotherton}, S.C.~{Gallagher}, Q.~{Ni}
  et~al., \emph{{On the Observational Difference between the Accretion
  Disk-Corona Connections among Super- and Sub-Eddington Accreting Active
  Galactic Nuclei}},
  \href{https://doi.org/10.3847/1538-4357/abe37f}{\emph{\apj} {\bfseries 910}
  (2021) 103} [\href{https://arxiv.org/abs/2102.02832}{{\ttfamily
  2102.02832}}].

\bibitem{Pu2020}
X.~{Pu}, B.~{Luo}, W.N.~{Brandt}, J.D.~{Timlin}, H.~{Liu}, Q.~{Ni} et~al.,
  \emph{{On the Fraction of X-Ray-weak Quasars from the Sloan Digital Sky
  Survey}}, \href{https://doi.org/10.3847/1538-4357/abacc5}{\emph{\apj}
  {\bfseries 900} (2020) 141}
  [\href{https://arxiv.org/abs/2008.02277}{{\ttfamily 2008.02277}}].

\bibitem{BorosonGreen1992}
T.A.~{Boroson} and R.F.~{Green}, \emph{{The Emission-Line Properties of
  Low-Redshift Quasi-stellar Objects}},
  \href{https://doi.org/10.1086/191661}{\emph{\apjs} {\bfseries 80} (1992)
  109}.

\bibitem{Wang1996}
T.~{Wang}, W.~{Brinkmann} and J.~{Bergeron}, \emph{{X-ray properties of active
  galactic nuclei with optical FeII emission.}}, {\emph{\aap} {\bfseries 309}
  (1996) 81}.

\bibitem{Richards2011}
G.T.~{Richards}, N.E.~{Kruczek}, S.C.~{Gallagher}, P.B.~{Hall}, P.C.~{Hewett},
  K.M.~{Leighly} et~al., \emph{{Unification of Luminous Type 1 Quasars through
  C IV Emission}},
  \href{https://doi.org/10.1088/0004-6256/141/5/167}{\emph{\aj} {\bfseries 141}
  (2011) 167} [\href{https://arxiv.org/abs/1011.2282}{{\ttfamily 1011.2282}}].

\bibitem{Timlin2021}
I.~{Timlin}, John~D., W.N.~{Brandt} and A.~{Laor}, \emph{{What controls the
  UV-to-X-ray continuum shape in quasars?}},
  \href{https://doi.org/10.1093/mnras/stab1217}{\emph{\mnras} {\bfseries 504}
  (2021) 5556} [\href{https://arxiv.org/abs/2104.13938}{{\ttfamily
  2104.13938}}].

\bibitem{Panessa2006}
F.~{Panessa}, L.~{Bassani}, M.~{Cappi}, M.~{Dadina}, X.~{Barcons},
  F.J.~{Carrera} et~al., \emph{{On the X-ray, optical emission line and black
  hole mass properties of local Seyfert galaxies}},
  \href{https://doi.org/10.1051/0004-6361:20064894}{\emph{\aap} {\bfseries 455}
  (2006) 173} [\href{https://arxiv.org/abs/astro-ph/0605236}{{\ttfamily
  astro-ph/0605236}}].

\bibitem{Ueda2015}
Y.~{Ueda}, Y.~{Hashimoto}, K.~{Ichikawa}, Y.~{Ishino}, A.Y.~{Kniazev},
  P.~{V{\"a}is{\"a}nen} et~al., \emph{{[O iii] {\ensuremath{\lambda}}5007 and
  X-Ray Properties of a Complete Sample of Hard X-Ray Selected AGNs in the
  Local Universe}},
  \href{https://doi.org/10.1088/0004-637X/815/1/1}{\emph{\apj} {\bfseries 815}
  (2015) 1} [\href{https://arxiv.org/abs/1510.03153}{{\ttfamily 1510.03153}}].

\bibitem{Melendez2008}
M.~{Mel{\'e}ndez}, S.B.~{Kraemer}, B.K.~{Armentrout}, R.P.~{Deo},
  D.M.~{Crenshaw}, H.R.~{Schmitt} et~al., \emph{{New Indicators for AGN Power:
  The Correlation between [O IV] 25.89 {\ensuremath{\mu}}m and Hard X-Ray
  Luminosity for Nearby Seyfert Galaxies}},
  \href{https://doi.org/10.1086/588807}{\emph{\apj} {\bfseries 682} (2008) 94}
  [\href{https://arxiv.org/abs/0804.1147}{{\ttfamily 0804.1147}}].

\bibitem{Kakkad2020}
D.~{Kakkad}, V.~{Mainieri}, G.~{Vietri}, S.~{Carniani}, C.M.~{Harrison},
  M.~{Perna} et~al., \emph{{SUPER. II. Spatially resolved ionised gas
  kinematics and scaling relations in z {\ensuremath{\sim}} 2 AGN host
  galaxies}}, \href{https://doi.org/10.1051/0004-6361/202038551}{\emph{\aap}
  {\bfseries 642} (2020) A147}
  [\href{https://arxiv.org/abs/2008.01728}{{\ttfamily 2008.01728}}].

\bibitem{Mulchaey1994}
J.S.~{Mulchaey}, A.~{Koratkar}, M.J.~{Ward}, A.S.~{Wilson}, M.~{Whittle},
  R.R.J.~{Antonucci} et~al., \emph{{Multiwavelength Tests of the Dusty Torus
  Model for Seyfert Galaxies}},
  \href{https://doi.org/10.1086/174933}{\emph{\apj} {\bfseries 436} (1994)
  586}.

\bibitem{Heckman2005}
T.M.~{Heckman}, A.~{Ptak}, A.~{Hornschemeier} and G.~{Kauffmann}, \emph{{The
  Relationship of Hard X-Ray and Optical Line Emission in Low-Redshift Active
  Galactic Nuclei}}, \href{https://doi.org/10.1086/491665}{\emph{\apj}
  {\bfseries 634} (2005) 161}
  [\href{https://arxiv.org/abs/astro-ph/0507674}{{\ttfamily
  astro-ph/0507674}}].

\bibitem{Pennell2017}
A.~{Pennell}, J.C.~{Runnoe} and M.S.~{Brotherton}, \emph{{Updating quasar
  bolometric luminosity corrections - III. [O III] bolometric corrections}},
  \href{https://doi.org/10.1093/mnras/stx556}{\emph{\mnras} {\bfseries 468}
  (2017) 1433} [\href{https://arxiv.org/abs/1703.03431}{{\ttfamily
  1703.03431}}].

\bibitem{Runnoe2012}
J.C.~{Runnoe}, M.S.~{Brotherton} and Z.~{Shang}, \emph{{Updating quasar
  bolometric luminosity corrections}},
  \href{https://doi.org/10.1111/j.1365-2966.2012.20620.x}{\emph{\mnras}
  {\bfseries 422} (2012) 478}
  [\href{https://arxiv.org/abs/1201.5155}{{\ttfamily 1201.5155}}].

\end{thebibliography}\endgroup

\bigskip
\bigskip
\noindent {\bf DISCUSSION}

\bigskip
\noindent {\bf MARIA GIOVANNA DAINOTTI's Comment:} The consistency of the slope in the Risaliti-Lusso relation is due to the fact that when you correct for the luminosity evolution then the effect of these correction gives the constancy of the slope. In the application of cosmology this correction must be applied in order to avoid biases in the cosmological parameters. A set-up methodology has been fully developed in the following four published papers: 1. Dainotti et al. 2023, arXiv230519668D, 2. Bargiacchi, Dainotti et al. 2023, MNRAS, 521, 3909B 3. Lenart, Bargiacchi, Dainotti et al. 2023 ApJS, 264, 46L 4. Dainotti et al. 2022, ApJ, 931, 106D

\end{document}